\documentclass[preprint,aps,nofootinbib]{revtex4}
\usepackage{amsfonts,amsmath,amssymb,amsthm}
\usepackage{latexsym}
\usepackage{bbm,bm}
\usepackage{graphicx}

\newcommand{\ket}[1]{\lvert #1 \rangle}
\newcommand{\bra}[1]{\langle #1 \lvert}
\newcommand{\beq}{\begin{equation}}
\newcommand{\eeq}{\end{equation}}
\newcommand{\beqs}{\begin{eqnarray}}
\newcommand{\eeqs}{\end{eqnarray}}

\begin{document}

\title{Time dependence of the position-momentum and position-velocity uncertainties in gapped graphene}
\author{Eylee Jung$^1$, Kwang S. Kim$^1$, and DaeKil Park$^{2,3}$}

\affiliation{$^1$Center for Superfunctional Materials,
                 Department of Chemistry,
                 Pohang University of Science and Technology, 
                 San 31, Hyojadong, Namgu, Pohang 790-784, Korea \\
             $^2$Department of Physics, Kyungnam University, ChangWon,
                  631-701, Korea                                        \\
             $^3$Department of Electronic Engineering, Kyungnam University, ChangWon,
              631-701, Korea                     
             }

\begin{abstract}
We examine the time-dependence of the position-momentum and position-velocity uncertainties in the 
monolayer gapped graphene. The effect of the energy gap to the uncertainties is shown to appear via the Compton-like wavelength $\lambda_c$. 
The uncertainties in the graphene are mainly contributed by two phenomena, spreading and zitterbewegung. While the former determines the uncertainties 
in the long-range of time, the latter gives high oscillation to the uncertainties in the short-range of time. The uncertainties in the 
graphene are compared with the corresponding values for the usual free Hamiltonian $\hat{H}_{free} = (p_1^2 + p_2^2) / 2 M$. It is shown that 
the uncertainties can be under control within the quantum mechanical laws if one can choose the gap parameter $\lambda_c$ freely.
\end{abstract}


\maketitle

\section{Introduction}

After success for fabricating the monolayer or few layer graphene\cite{fabrication}, there are a lot of activities for researching into the 
various properties of graphene\cite{review}. This is mainly due to the fact that the low-energy electrons in graphene have unusual electronic properties. 

Long ago it was predicted by Wallace\cite{wallace} that the electron located near the hexagonal vertices of the Brillouin zone exhibits a
linear dispersion 
relation and $40$ years later Semenoff\cite{semenoff} showed that the low-energy dynamics of the corresponding electron is governed by massless Dirac equation even in the non-relativistic regime. Thus, the fabrication of the monolayer graphene opens a possibility to test various predictions of 
quantum electrodynamics (QED) by making use of condensed matter experiment. However, this does not mean that all phenomena QED predicted can 
be realized in the graphene-based experiment because the light velocity $c$ in QED should be replaced by the Fermi velocity $v_F \sim c / 300$. It 
results in the large fine-structure constant $\alpha \sim 2$. This fact implies that only non-perturbative characters of the planar QED can 
be realized in the graphene experiment. Recently, there have been many efforts on this connection\cite{connection}. 

Among many phenomena arising in the planar QED most interesting issue, at least for us, is the spin-1/2 Aharonov-Bohm (AB)\cite{ab1959} or 
Aharonov-Bohm-Coulomb (ABC) problem, which was extensively discussed about two decades ago\cite{spin-ab} because the same problem appeared 
in the context of anyonic and cosmic string theories\cite{cosmic}. The most important issue in this problem is how to treat the
$\delta$-like singular potential generated by an interaction between particle's spin and thin magnetic flux tube. Recently, similar AB and related problems were discussed theoretically\cite{ab-graphene-th} and experimentally\cite{ab-graphene-exp} in the branch of graphene physics. Another closely
related issue in the graphene is Coulomb impurity problem\cite{coulomb}. The interesting fact in this case is that depending on the charge of 
impurity there are two regions, subcritical and supercritical, in which the effects of impurity are completely different. Similar phenomenon in 
QED was discussed long ago in Ref.\cite{zeldo}. 

Other unobserved interesting phenomena QED predicts are Klein paradox and zitterbewegung.
The Klein paradox\cite{klein}-counterintuitive barrier penetration in the relativistic setting-was re-examined in Ref.\cite{paradox1}. 
Authors in Ref.\cite{paradox1} argued that the Klein paradox can be realized using electrostatic barriers in single- and bi-layer graphene. Few years 
later it was reported that the Klein tunneling was observed by measuring the quantum conductance oscillation and phase shift pattern in the extremely 
narrow graphene\cite{paradox2}. The zitterbewegung (ZB)\cite{zitter1}-the trembling motion arising due to the interference between positive and negative energy states- was also investigated recently in the graphene  without\cite{zitter2} and with\cite{zitter3} external magnetic field. The 
effect of zitterbewegung for other models is also discussed recently.\cite{zitter4}

Besides a connection between graphene and QED much attractive attention is paid to the graphene as a new material for future technology. Most 
important application of graphene, at least for us, is a possibility for realization of quantum computer. Recently, many techniques are
used independently or cooperatively to realize the quantum computer. The typical techniques are optical ones, ion traps, NMR, quantum dots, 
and superconductors. Current status for the realization is summarized in detail in Ref. \cite{qcs}. Also, the graphene-based quantum 
computer is explored in Ref. \cite{qc1}.

In this paper we will examine the position-momentum and position-velocity uncertainties of the low-energy electrons in the 
monolayer gapped graphene when the initial wave packet is chosen as a general Gaussian wave packet. Since Gaussian wave packet, in general,
contains both positive-energy and negative-energy spectra, the expectation values of the physical quantities should be 
contributed by spreading and zitterbewegung phenomena. Thus, it is of interest to examine the effect of the gap parameter in the 
expectation values of various quantities and uncertainties. We will show in this paper that the position-momentum and position-velocity
uncertainties can be under control within the quantum mechanical laws if the gap parameter can be chosen freely. 

Although this controllability of the uncertainties is an interesting fact from the aspect of purely theoretical ground, it is also important
from the aspect of quantum computer. Quantum computer\cite{qcs} is a machine, which performs quantum computational 
processes by making use of the quantum mechanical laws. So far many quantum information processes are developed such as quantum 
teleportation\cite{teleport}, factoring algorithm\cite{factoring}, and search algorithm\cite{search}. All quantum information processes
consist of three stages: preparation of initial states at initial stage, time evolution of quantum states via various unitary gates at  
intermediate stage, and quantum measurements at final stages. If uncertainties, therefore, are large at the final stage, the quantum 
measurement can generate fatal errors in the computing processes. For this reason it is important to reduce the uncertainties
as much as possible at the final stages.

This paper is organized as follows. In section II we examine the position-momentum uncertainties in the gapped graphene. It is shown that the 
uncertainties are contributed by the spreading and ZB effects of the given wave packet. The uncertainties in the gapped graphene are 
compared with the corresponding quantities of the $2d$ free Hamiltonian system. In section III we discuss on the position-velocity uncertainties 
in the gapped graphene. Unlike the position uncertainties the velocity uncertainties are shown to be contributed by sorely ZB effect of the wave
packet. This fact implies that the $t \rightarrow \infty$ limit of the velocity uncertainties coincides with the Fermi velocity $v_F$ regardless of 
the choice of the packet. In section IV a brief conclusion is given.

\section{position-momentum uncertainty}

In this section we examine the position-momentum uncertainty in the gapped graphene.
The appropriate Hamiltonian for the low-energy electron near the Dirac point is given by 
\begin{eqnarray}
\label{hamil-2-1}
\hat{H}_M = v_F \left(        \begin{array}{cc}
                     M v_F  &  p_1 - i p_2          \\
                     p_1 + i p_2  &  -M v_F
                            \end{array}            \right)
\end{eqnarray}
where $v_F \sim c / 300$ is a Fermi velocity and $M$ is a gap parameter generated via some dynamical 
and technical reasons. Theoretically, most popular mechanism which generates the gap is a chiral symmetry breaking\cite{dynamical3}. This 
mechanism is similar to dynamical breaking\cite{DGSB}, which was studied deeply in gauge theories. The bandgap can be generated 
by breaking the sublattice symmetry. This case was experimentally realized by choosing the substrate appropriately\cite{dynamical1}. 
In addition, the gap is also generated in graphene nanoribbon\cite{dynamical2}. Both cases are taken into account in Hamiltonian (\ref{hamil-2-1}).
Although monolayer graphene itself does not have a gap, the bandgap is naturally generated in bilayer graphene\cite{bilayer}. However, we 
cannot use the Hamiltonian (\ref{hamil-2-1}) to explore the effect of gap in the bilayer graphene due to non-trivial structure of the gap
in the bilayer system. From the terminology of relativistic field theories this gap parameter $M$ is a mass term of the Dirac fermion.

The position operator $\hat{x} (t)$ in the Heisenberg picture can be expressed by $2 \times 2$ matrix from 
$\hat{x} (t) = \exp(i \hat{H}_M t / \hbar) \hat{x} (0) \exp(-i \hat{H}_M t / \hbar)$. Explicit calculation shows 
\begin{eqnarray}
\label{position1}
\hat{x} (t) = \hat{x} (0) +         \left(       \begin{array}{cc}
                                           \hat{\Sigma} (p)  &  \hat{\sigma}_1 (p) + i \hat{\sigma}_2 (p)        \\
                                           \hat{\sigma}_1 (p) - i \hat{\sigma}_2 (p) &  -\hat{\Sigma} (p)
                                                  \end{array}                                      \right),
\end{eqnarray}
where
\begin{eqnarray}
\label{position2}
& &\hat{\Sigma} (p) = \frac{\hbar}{{\bm p}^2 + (M v_F)^2} 
\left[ p_2 \sin^2 \theta_M + \frac{(M v_F) p_1}{\sqrt{{\bm p}^2 + (M v_F)^2}} \left(\theta_M - \sin \theta_M \cos \theta_M \right) \right]
                                                                                              \\    \nonumber
& &\hat{\sigma}_1 (p) = \frac{\hbar}{[{\bm p}^2 + (M v_F)^2]^{3/2}} 
\left[\theta_M p_1^2 + \sin \theta_M \cos \theta_M \left\{p_2^2 + (M v_F)^2 \right\} \right]           \\    \nonumber
& &\hat{\sigma}_2 (p) = \frac{\hbar}{[{\bm p}^2 + (M v_F)^2]^{3/2}} 
\left[ p_1 p_2 \left(\sin \theta_M \cos \theta_M - \theta_M \right) + (M v_F) \sqrt{{\bm p}^2 + (M v_F)^2} \sin^2 \theta_M \right]
\end{eqnarray}
and $\theta_M = (v_F t / \hbar) \sqrt{{\bm p}^2 + (M v_F)^2}$. Each operator in Eq.(\ref{position2}) consists of two kinds, one of which is 
responsible for ZB phenomena and the other is responsible for the spreading of wave packet. 

In order to examine the uncertainty relations we should introduce a wave packet. In this paper we introduce a usual two-dimensional 
Gaussian wave packet
\begin{eqnarray}
\label{packet}
|\psi (x, y: 0)\rangle = \frac{d}{2 \pi \sqrt{\pi}} \int d^2 {\bm k} \exp \left[-\frac{d^2}{2} (k_x - \alpha)^2 - \frac{d^2}{2} (k_y - \beta)^2\right]
e^{i {\bm k} \cdot {\bm r}}   
\left(     \begin{array}{c}
                a            \\
                b
          \end{array}             \right),
\end{eqnarray}
where real parameters $a$ and $b$ satisfy $a^2 + b^2 = 1$. It is easy to show that $|\psi (x, y: 0)\rangle$ can be decomposed as 
\begin{equation}
\label{decompose1}
|\psi (x, y: 0)\rangle = |\psi^p (x, y: 0)\rangle + |\psi^n (x, y: 0)\rangle,
\end{equation}
where $|\psi^p (x, y: 0)\rangle$ and $|\psi^n (x, y: 0)\rangle$ are the positive-energy and negative-energy components of 
$|\psi (x, y: 0)\rangle$, respectively. Using Hamiltonian $\hat{H}_M$ it is easy to derive these components and the explicit 
expressions are given by
\begin{eqnarray}
\label{decompose2}
& &|\psi^p (x, y: 0)\rangle = \frac{d}{4 \pi \sqrt{\pi}} \int d^2 {\bm k} \exp \left[-\frac{d^2}{2} (k_x - \alpha)^2 - \frac{d^2}{2} (k_y - \beta)^2\right]
e^{i {\bm k} \cdot {\bm r}}                                                          \\   \nonumber
& &  \hspace{3.0cm} \times                    
\frac{a k_+ + b (\sqrt{{\bm k}^2 + \lambda_c^{-2}} - \lambda_c^{-1})}{k_+ \sqrt{{\bm k}^2 + \lambda_c^{-2}}} 
\left(     \begin{array}{c}
                \sqrt{{\bm k}^2 + \lambda_c^{-2}} +  \lambda_c^{-1}           \\
                k_+
          \end{array}             \right)                                                               \\   \nonumber
& &|\psi^n (x, y: 0)\rangle = \frac{d}{4 \pi \sqrt{\pi}} \int d^2 {\bm k} \exp \left[-\frac{d^2}{2} (k_x - \alpha)^2 - \frac{d^2}{2} (k_y - \beta)^2\right]
e^{i {\bm k} \cdot {\bm r}}                                                                            \\   \nonumber
& &  \hspace{3.0cm} \times  
\frac{a k_+ - b (\sqrt{{\bm k}^2 + \lambda_c^{-2}} + \lambda_c^{-1})}{k_+ \sqrt{{\bm k}^2 + \lambda_c^{-2}}} 
\left(     \begin{array}{c}
                \sqrt{{\bm k}^2 + \lambda_c^{-2}} -  \lambda_c^{-1}           \\
                -k_+
          \end{array}             \right).
\end{eqnarray}
In Eq. (\ref{decompose2}) $k_{\pm} = k_x \pm i k_y$ and $\lambda_c = \hbar / (M v_F)$. The parameter $\lambda_c$ is a familiar quantity.
In fact, this is a Compton wavelength if the Fermi velocity $v_F$ is replaced with light velocity $c$. In this paper we will call 
$\lambda_c$ as Compton wavelength. Thus, the intensity for the positive-energy and negative-energy components are
\begin{eqnarray}
\label{decompose3}
& &P_+ \equiv \langle \psi^p (x, y: 0) | \psi^p (x, y: 0) \rangle = \frac{1}{2} + \Delta P             \\   \nonumber
& &P_- \equiv \langle \psi^n (x, y: 0) | \psi^n (x, y: 0) \rangle = \frac{1}{2} - \Delta P = 1 - P_+,
\end{eqnarray}
where
\begin{equation}
\label{decompose4}
\Delta P = \frac{d^2}{2 \pi} \int d^2 {\bm k} \exp \left[-d^2 (k_x - \alpha)^2 - d^ (k_y - \beta)^2\right]
\frac{\lambda_c^{-1} (a^2 - b^2) + 2 a b k_x}{\sqrt{{\bm k}^2 + \lambda_c^{-2}}}.
\end{equation}
If, therefore, $\alpha = 0$ with $a = b = 1 / \sqrt{2}$, we get $P_+ = P_- = 1/2$. In this case the expectation values of various 
operators are summarized in Appendix A. For arbitrary $\alpha$ and $\beta$, however, 
$P_{\pm}$ should be computed numerically. Since $|\psi (x, y: 0)\rangle$ has both positive-energy and negative-energy components, the expectation
value of various physical quantities should exhibit the trembling behavior due to the interference of these components as discussed in 
Ref.\cite{zitter1,zitter2,zitter3,zitter4}.

Using Eq.(\ref{position1}) and Eq.(\ref{packet}) it is straightforward to show
\begin{equation}
\label{position3}
\langle x \rangle (t) \equiv \bra{\psi (x, y: 0)} \hat{x} (t) \ket{\psi (x, y: 0)}
= \frac{d^2}{\pi} \int d^2 {\bm k} \exp \left[-d^2  (k_x - \alpha)^2 - d^2 (k_y - \beta)^2\right] \left(X_S + X_{ZB} \right),
\end{equation}
where
\begin{eqnarray}
\label{position4}
& &X_S = \frac{(v_F t)}{{\bm k}^2 + \lambda_c^{-2}} \left[(a^2 - b^2) \lambda_c^{-1} k_x + 2 a b k_x^2 \right]     \\    \nonumber
& &X_{ZB} = \frac{a^2 - b^2}{{\bm k}^2 + \lambda_c^{-2}} \left[ k_y \sin^2 \theta - \frac{\lambda_c^{-1} k_x}{\sqrt{{\bm k}^2 + \lambda_c^{-2}}}
\sin \theta \cos \theta \right] + \frac{2 a b}{({\bm k}^2 + \lambda_c^{-2})^{3/2}} \sin \theta \cos \theta (k_y^2 + \lambda_c^{-2})
\end{eqnarray}
and, $\theta = (v_F t) \sqrt{{\bm k}^2 + \lambda_c^{-2}}$. 
As remarked before $X_S$ and $X_{ZB}$ are responsible for the spreading and trembling motion in the 
time evolution of the packet, respectively. It is worthwhile noting that the ${\bm k}$-integration in Eq.(\ref{position3}) can be performed 
explicitly by making use of the binomial expansion. Finally, then, $\langle x \rangle (t)$ is represented in terms of the Hermite polynomials.
Instead of integral representation, however, $\langle x \rangle (t)$ has triple summations. The explicit expressions in terms of the Hermite 
polynomials for various expectation values derived in this paper are summarized in Appendix B.

Similar calculation procedure derives $\langle y \rangle (t)$ as 
\begin{equation}
\label{position5}
\langle y \rangle (t) \equiv \bra{\psi (x, y: 0)} \hat{y} (t) \ket{\psi (x, y: 0)}
= \frac{d^2}{\pi} \int d^2 {\bm k} \exp \left[-d^2  (k_x - \alpha)^2 - d^2 (k_y - \beta)^2\right] \left(Y_S + Y_{ZB} \right),
\end{equation}
where
\begin{eqnarray}
\label{position6}
& &Y_S = \frac{(v_F t)}{{\bm k}^2 + \lambda_c^{-2}} \left[(a^2 - b^2) \lambda_c^{-1} k_y + 2 a b k_x k_y \right]     \\    \nonumber
& &Y_{ZB} = \frac{\sin^2 \theta}{{\bm k}^2 + \lambda_c^{-2}} \left[ -(a^2 - b^2) k_x + 2 a b \lambda_c^{-1} \right] 
- \frac{\sin \theta \cos \theta}{({\bm k}^2 + \lambda_c^{-2})^{3/2}} \left[ (a^2 - b^2) \lambda_c^{-1} k_y + 2 a b k_x k_y \right].
\end{eqnarray}
Of course, $Y_S$ and $Y_{ZB}$ represent the spreading and ZB motion of the wave packet in $y$-direction.

In order to confirm the validity of our calculation we consider the case of zero gap ($\lambda_c^{-1} \rightarrow 0$), which was considered in
Ref.\cite{zitter2}. For simplicity, we choose $\alpha = 0$, $a=1$, and $b=0$. Then, $Y_S = 0$ and 
$Y_{ZB} = -\sin^2 \theta k_x / {\bf k}^2$, which makes $\langle y \rangle (t) = 0$ due to $k_x$-integration. In this case we also get 
$X_S = 0$ and $X_{ZB} = \sin^2 \theta k_y / {\bf k}^2$. Using $\int_0^{2\pi} d\theta \sin \theta e^{a \sin \theta} = 2\pi I_1 (a)$, 
where $I_{\nu} (z)$ is a modified Bessel function, one can show directly
\begin{equation}
\label{gapless1}
\langle x \rangle (t) = \frac{1}{2\beta} \left(1 - e^{-\beta^2 d^2} \right) - d e^{-\beta^2 d^2} 
\int_{0}^{\infty} dq e^{-q^2} \cos \left(\frac{2 v_F t}{d} q \right) I_1 (2 \beta d q),
\end{equation}
which exactly coincides with the second reference of Ref.\cite{zitter2}.

Before we explore the uncertainty properties it is interesting to examine the limiting behaviors of $\langle x \rangle (t)$
and $\langle y \rangle (t)$. In the $t \rightarrow 0$ limit some combinations of the spreading and trembling motion become dominant 
and the limiting behaviors reduce to
\begin{eqnarray}
\label{limit1}
& &\lim_{t \rightarrow 0} \langle x \rangle (t) = 2 a b (v_F t) + O \left((v_F t)^2 \right)                 \\    \nonumber
& &\lim_{t \rightarrow 0} \langle y \rangle (t) = (v_F t)^2 \left[-(a^2 - b^2) \alpha + 2 a b \lambda_c^{-1} \right] + 
O \left((v_F t)^3 \right).
\end{eqnarray}
It is interesting to note that the $t \rightarrow 0$ limiting behaviors of $\langle x \rangle (t)$ and $\langle y \rangle (t)$ are 
completely different because their orders of $v_F t$ are different from each other. Furthermore, the dominant terms of 
$\langle x \rangle (t)$ come from the off-diagonal components of $\hat{x} (t)$ while those of $\langle y \rangle (t)$ are 
contributed from all components of $\hat{y} (t)$. In the $t \rightarrow \infty$ limit the dominant terms in 
$\langle x \rangle (t)$ and $\langle y \rangle (t)$ are contributed from spreading terms and their expressions are 
\begin{eqnarray}
\label{limit2}
& &\lim_{t \rightarrow \infty} \langle x \rangle (t) = \frac{d^2 (v_F t)}{\pi} \left[ (a^2 - b^2) \lambda_c^{-1} J_{1,0}
 + 2 a b J_{2,0} \right]                             \\    \nonumber
& &\lim_{t \rightarrow \infty} \langle y \rangle (t) = \frac{d^2 (v_F t)}{\pi} \left[ (a^2 - b^2) \lambda_c^{-1} J_{0,1}
 + 2 a b J_{1,1} \right],
\end{eqnarray}
where
\begin{equation}
\label{limit3}
J_{m,n} \equiv  \int d^2 {\bm k} \exp \left[-d^2  (k_x - \alpha)^2 - d^2 (k_y - \beta)^2\right]
\frac{k_x^m  k_y^n}{{\bm k}^2 + \lambda_c^{-2}}.
\end{equation}

In order to examine the position uncertainty $\Delta x (t)$ we should derive $\hat{x}^2 (t)$, which reduces to 
\begin{equation}
\label{psquare1}
\hat{x}^2 (t) = \left[ \hat{x}^2 (0) + \hat{\Sigma}^2 (p) + \hat{\sigma}_1^2 (p) + \hat{\sigma}_2^2 (p) \right] \openone + 
\left\{ \hat{x} (0), \hat{x} (t) - \hat{x} (0) \right\},
\end{equation}
where $\left\{ A, B \right\} \equiv A B + B A$. Since it is straightforward to show 
$\bra{\psi (x, y: 0)}  \left\{ \hat{x} (0), \hat{Z} (p) \right\} \ket{\psi (x, y: 0)} = 0$ with $\hat{Z} = \hat{\Sigma}$, $\hat{\sigma}_1$, or 
$\hat{\sigma}_2$, one can show directly
\begin{equation}
\label{psquare2}
\langle x^2 \rangle (t) = \frac{d^2}{2} + \frac{d^2}{\pi} 
\int d^2 {\bm k} \exp \left[-d^2  (k_x - \alpha)^2 - d^2 (k_y - \beta)^2\right] \left(\tilde{X}_S + \tilde{X}_{ZB} \right),                                
\end{equation}
where 
\begin{equation}
\label{psquare3}
\tilde{X}_S = (v_F t)^2 \frac{k_x^2}{{\bm k}^2 + \lambda_c^{-2}}     \hspace{2.0cm}
\tilde{X}_{ZB} = \sin^2 \theta \frac{k_y^2 + \lambda_c^{-2}}{({\bm k}^2 + \lambda_c^{-2})^2}.
\end{equation}
Similar calculation shows 
\begin{equation}
\label{psquare4}
\langle y^2 \rangle (t) = \frac{d^2}{2} + \frac{d^2}{\pi} 
\int d^2 {\bm k} \exp \left[-d^2  (k_x - \alpha)^2 - d^2 (k_y - \beta)^2\right] \left(\tilde{Y}_S + \tilde{Y}_{ZB} \right),                                
\end{equation}
where $\tilde{Y}_S$ and $\tilde{Y}_{ZB}$ are obtained from $\tilde{X}_S$ and $\tilde{X}_{ZB}$ by interchanging $k_x$ with $k_y$. 

For the case of zero gap ($\lambda_c^{-1} \rightarrow o$) with $\alpha = 0$, $a = 1$, and $b=0$ one can show straightforwardly
\begin{eqnarray}
\label{gapless2}
& &\langle x^2 \rangle(t) = \frac{d^2}{2} + \frac{(v_F t)^2}{2 \beta^2 d^2} \left(1 - e^{-\beta^2 d^2} \right)  \\   \nonumber
& & \hspace{1.5cm}
+ d^2 e^{-\beta^2 d^2} \int_0^{\infty} \frac{dq}{q^2} e^{-q^2}
\left[ 1 - \cos \left( \frac{2 v_F t}{d} q \right) \right] 
\left[ q I_0 (2 \beta d q) - \frac{1}{2 \beta d} I_1 (2 \beta d q) \right]                                       \\    \nonumber
& &\langle y^2 \rangle(t) = \frac{d^2}{2} + (v_F t)^2 
\left[ e^{-\beta^2 d^2 / 2} \left( \sin \frac{\beta^2 d^2}{2} + \cos \frac{\beta^2 d^2}{2} \right) - \frac{1}{2 \beta^2 d^2} 
\left(1 - e^{-\beta^2 d^2} \right)  \right]                                                                      \\   \nonumber
& & \hspace{1.5cm}
+ \frac{d}{2 \beta} e^{-\beta^2 d^2} \int_0^{\infty}  \frac{dq}{q^2} e^{-q^2} \left[ 1 - \cos \left( \frac{2 v_F t}{d} q \right) \right]
I_1 (2 \beta d q).
\end{eqnarray}
Eq. (\ref{gapless1}) and Eq. (\ref{gapless2}) can be used to compute the uncertainties $\Delta x$ and $\Delta y$ for the case of zero gap.

While in the $t \rightarrow 0$ limit $\langle x^2 \rangle (t)$ and $\langle y^2 \rangle (t)$ exhibit similar behavior as 
\begin{equation}
\label{limit4}
\lim_{t \rightarrow 0} \langle x^2 \rangle (t) = \lim_{t \rightarrow 0} \langle y^2 \rangle (t) = \frac{d^2}{2} + (v_F t)^2 + 
O \left( (v_F t)^3 \right),
\end{equation}
and the $t \rightarrow \infty$ limits of  $\langle x^2 \rangle (t)$ and $\langle y^2 \rangle (t)$ reduce to 
\begin{equation}
\label{limit5}
\lim_{t \rightarrow \infty} \langle x^2 \rangle (t) = \frac{d^2}{2} + \frac{d^2}{\pi} (v_F t)^2 J_{2,0} 
\hspace{1.0cm}                                                                                        
\lim_{t \rightarrow \infty} \langle y^2 \rangle (t) = \frac{d^2}{2} + \frac{d^2}{\pi} (v_F t)^2 J_{0,2}.
\end{equation}

\begin{figure}[ht!]
\begin{center}
\includegraphics[height=6.2cm]{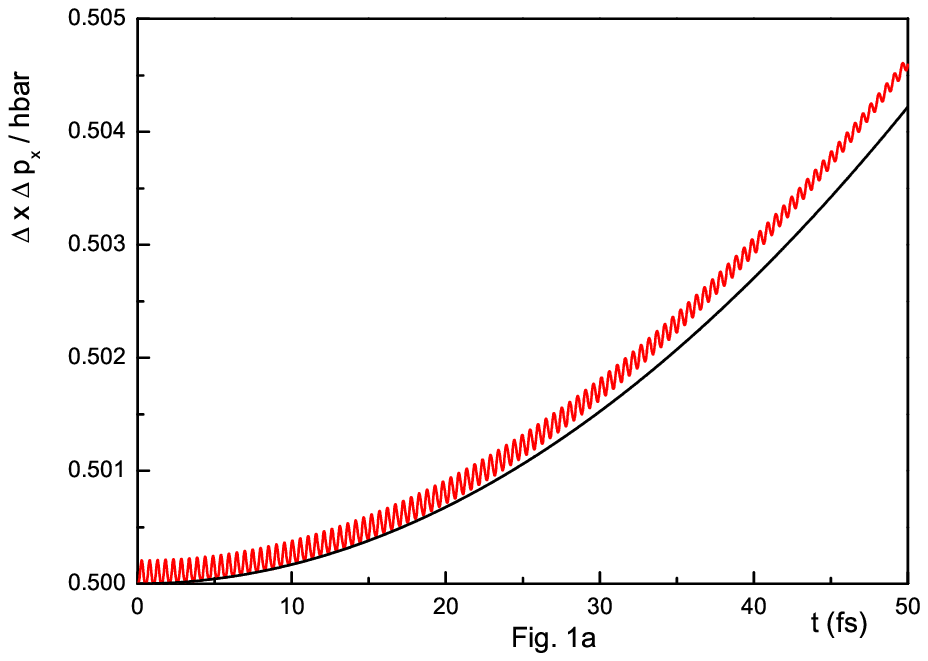}
\includegraphics[height=6.2cm]{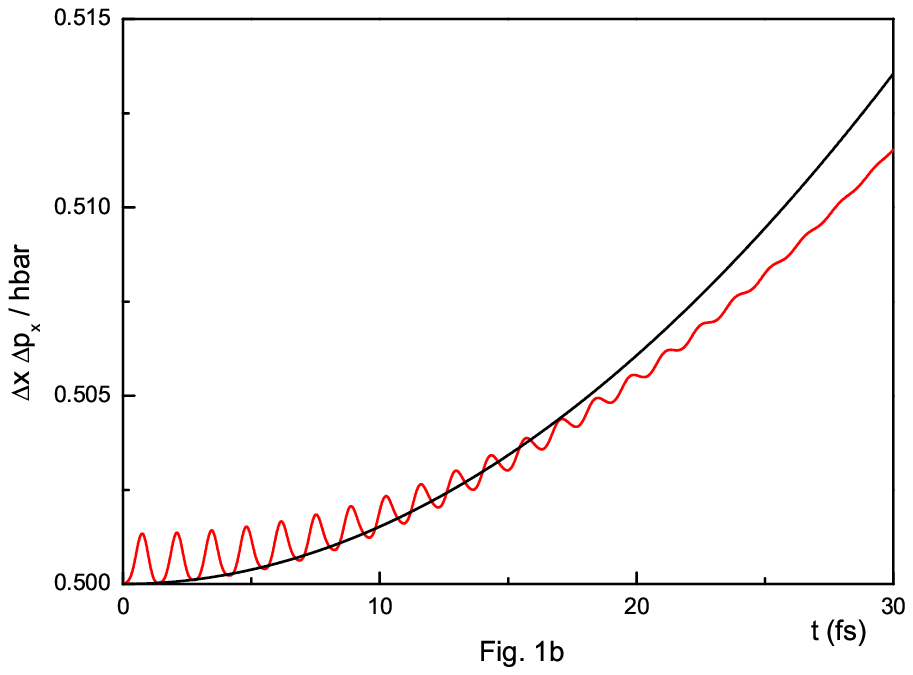}
\includegraphics[height=6.2cm]{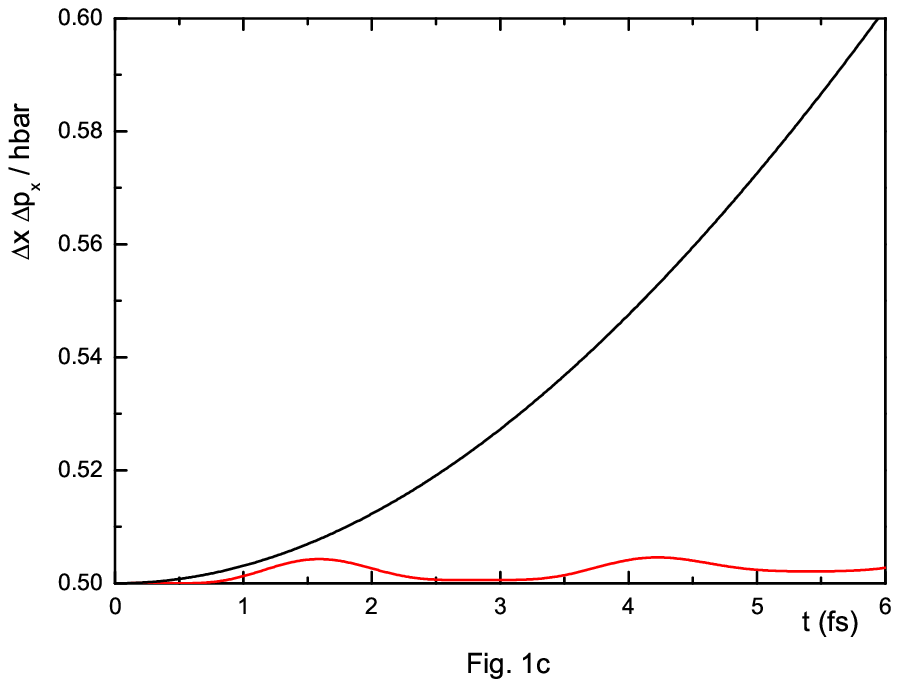}
\includegraphics[height=6.2cm]{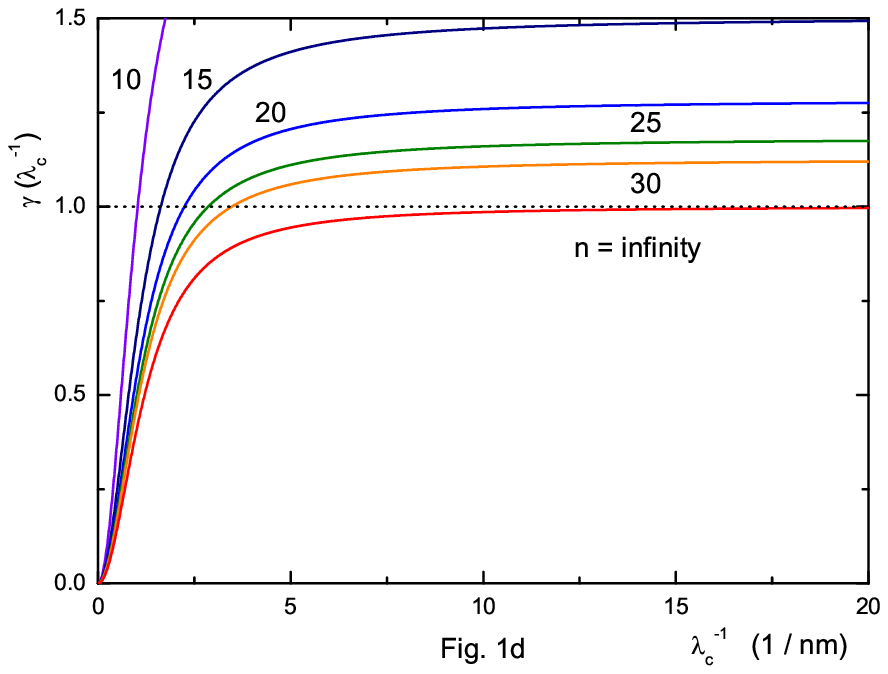}
\caption[fig1]{(Color online) The time-dependence of $\Delta x \Delta p_x / \hbar$ for 
$\lambda_c^{-1} = 6 (1 / nm)$ (a), $\lambda_c^{-1} = 2 (1 / nm)$ (b), 
and $\lambda_c^{-1} = 0.14 (1 / nm)$ (c). The black solid line for each figure is a corresponding value $(\Delta x \Delta p_x / \hbar)_{free}$ for 
the usual two-dimensional free Hamiltonian $\hat{H}_{free}$. As Fig. (a), (b), and (c) show, the uncertainty 
$\Delta x \Delta p_x$ in graphene is larger (or smaller) than $(\Delta x \Delta p_x)_{free}$ in the entire range of time when 
$\lambda_c^{-1} > \mu_2$ (or $\lambda_c^{-1} < \mu_1$). When $\mu_1 < \lambda_c^{-1} < \mu_2$, $\Delta x \Delta p_x$ is larger and smaller than
$(\Delta x \Delta p_x)_{free}$ at $t \rightarrow 0$ and $t \rightarrow \infty$ limits, respectively. (d) The critical value 
$\mu_2$ increases with decreasing $\alpha$, and eventually goes to $\infty$ at $\alpha = 0$.}
\end{center}
\end{figure}

Since it is easy to show $\Delta p_x = \Delta p_y = \hbar / \sqrt{2} d$, we plot the time-dependence of the dimensionless quantity 
$\Delta x \Delta p_x / \hbar$ in Fig. 1. In the figure we choose $a=0.9$, $d=8 (nm)$, $\alpha = 0.04 (1 / nm)$, and 
$\beta = 1.2 (1 / nm)$. We also choose the inverse of the Compton wave length as $6 (1/nm)$ (Fig. 1a), $2 (1 / nm)$ (Fig. 1b), and 
$0.14 (1/ nm)$ (Fig. 1 c). The black solid line in (a), (b), and (c) is $(\Delta x \Delta p_x / \hbar)_{free} = \sqrt{(1/2)^2 + (\lambda_c v_F t / 2 d^2)^2}$, 
which is a corresponding value for the usual non-relativistic free Hamiltonian $\hat{H}_{free} = (p_1^2 + p_2^2) / 2 M$. The unit of the 
time-axis is femto-second. 

As Fig. 1 represents, the uncertainty $\Delta x \Delta p_x$ has several distinct properties. First, it is contributed from both 
spreading and ZB motion of the wave packet. The spreading motion is dominated in the large scale of time. With increasing the inverse
Compton wavelength the overall increasing rate of $\Delta x \Delta p_x$ arising due to the spreading of the packet decreases drastically. 
This can be understood from the analogy of the relativistic field theories, that is, with increasing $M$ the relativistic theories approach
to the non-relativistic Galilean theories, where the uncertainty is minimized. In the small scale of time $\Delta x \Delta p_x$ 
oscillates rapidly due to the ZB effect. The amplitude of the oscillation increases with decreasing $\lambda_c^{-1}$. This is mainly due to 
the fact the the ZB effect is dominated when the energy gap $\Delta E$ between positive and negative energy spectra decreases. 
However, the frequency increases rapidly with increasing $\lambda_c^{-1}$ because of the famous formula $\omega = \Delta E / \hbar$.
When $\lambda_c^{-1}$ is larger than a critical value $\mu_2$, 
$\Delta x \Delta p_x$  becomes larger than $(\Delta x \Delta p_x)_{free}$ as Fig. 1a indicated. When, however, $\lambda_c^{-1}$ is smaller than 
a different critical value $\mu_1$, it is smaller than $(\Delta x \Delta p_x)_{free}$ as Fig. 1c shows. 
In the intermediate range of $\lambda_c^{-1}$ $\Delta x \Delta p_x$ is larger
and smaller than $(\Delta x \Delta p_x)_{free}$ in $t \rightarrow 0$ and $t \rightarrow \infty$ limits, respectively as Fig. 1b shows. 
Using Eq.(\ref{limit1}), (\ref{limit2}) and several other limiting values, one can derive the critical values $\mu_1$ explicitly, and 
$\mu_2$ implicitly as
\begin{equation}
\label{critical1}
\mu_1 = \frac{1}{\sqrt{2 d^2 (1 - 4 a^2 b^2)}},   \hspace{1.5cm}   \gamma (\lambda_c^{-1}) \bigg|_{\lambda_c^{-1} = \mu_2} = 1,
\end{equation}
where
\begin{equation}
\label{critical2}
\gamma (\lambda_c^{-1}) = \frac{2 \lambda_c^{-2} d^4}{\pi} \left[J_{2,0} - \frac{d^2}{\pi}\left\{ (a^2 - b^2) \lambda_c^{-1} J_{1,0} + 2 a b J_{2,0} \right\}^2 \right].
\end{equation}
The $\lambda_c^{-1}$-dependence of $\gamma (\lambda_c^{-1})$ is plotted in Fig. 1d when $a = 0.9$, $d = 8 (nm)$, $\alpha = 1.2 / n (1/nm)$, 
and $\beta = 1.2 (1/nm)$ for various $n$. As this figure indicates, the critical value $\mu_2$ increases with increasing $n$, and eventually 
$\mu_2 = \infty$ when $\alpha = 0$.

\begin{figure}[ht!]
\begin{center}
\includegraphics[height=6.2cm]{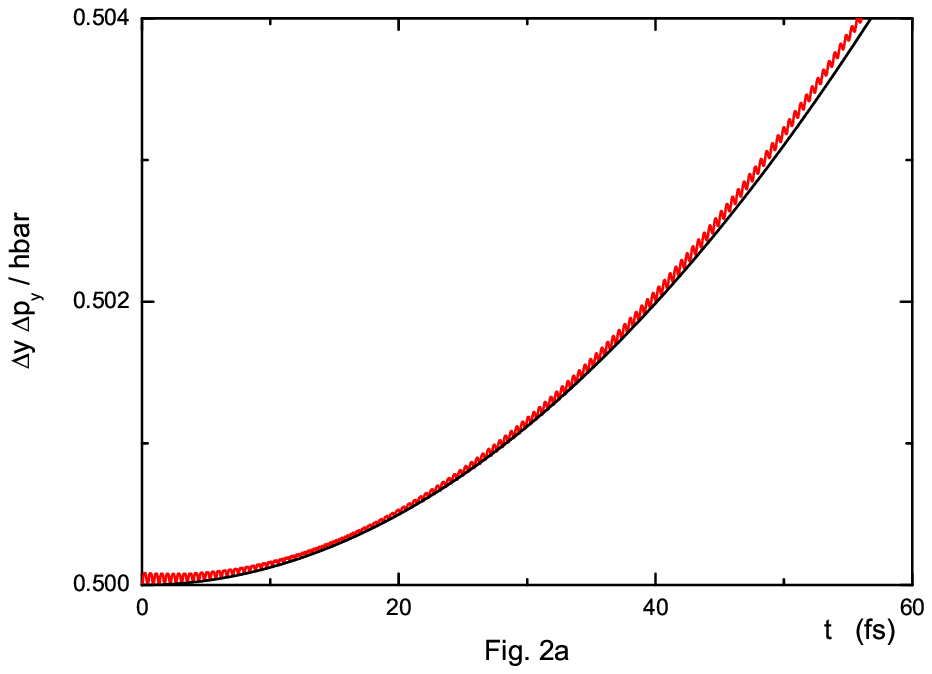}
\includegraphics[height=6.2cm]{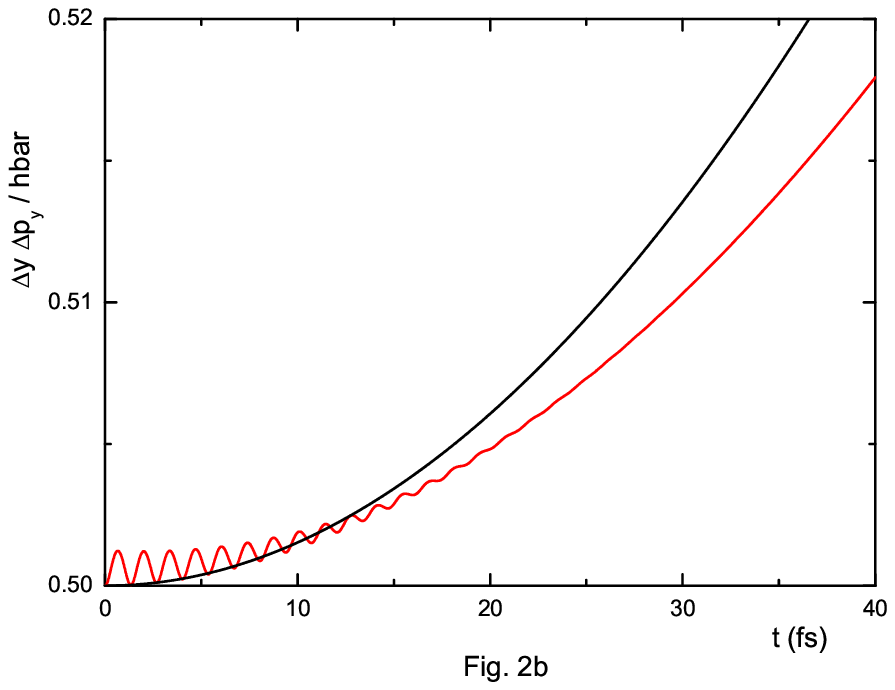}
\includegraphics[height=6.2cm]{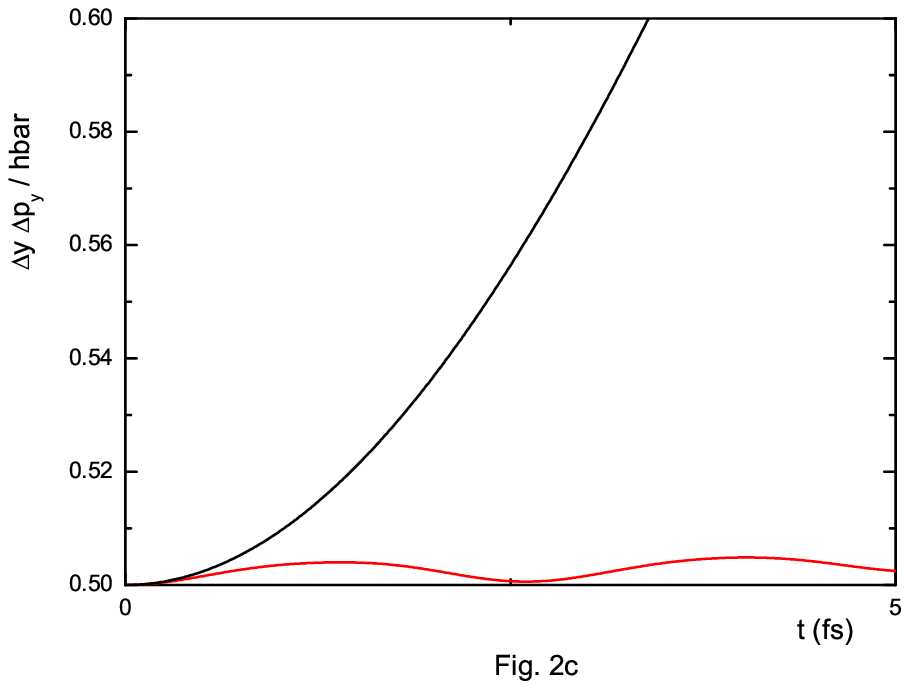}
\includegraphics[height=6.2cm]{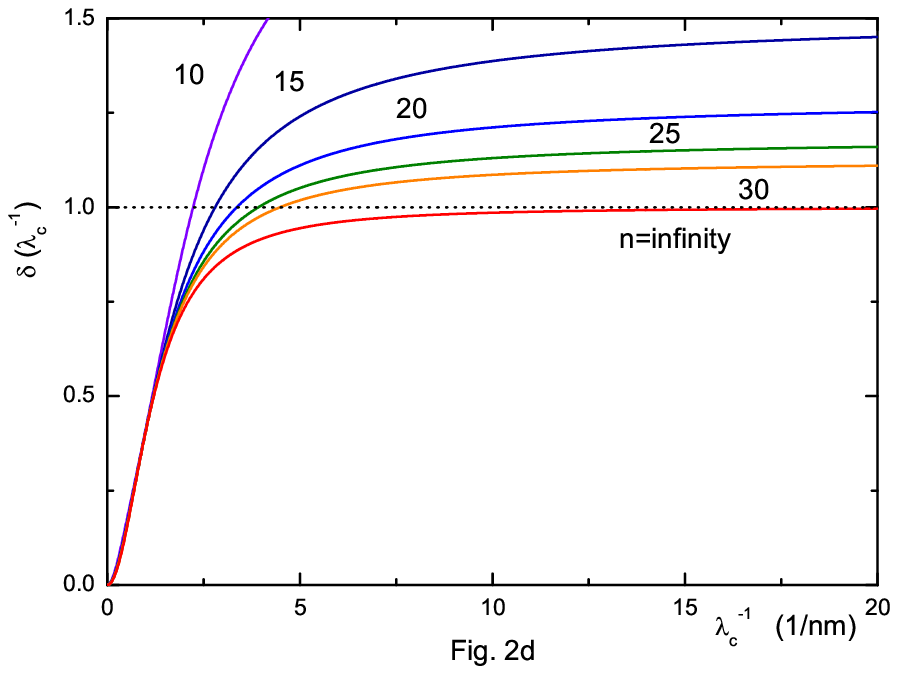}
\caption[fig2]{(Color online) The time-dependence of $\Delta y \Delta p_y / \hbar$ for 
$\lambda_c^{-1} =8 (1 / nm)$ (a), $\lambda_c^{-1} = 2 (1 / nm)$ (b), 
and $\lambda_c^{-1} = 0.04 (1 / nm)$ (c). The black solid line for each figure is a corresponding value $(\Delta y \Delta p_y / \hbar)_{free}$.  
As Fig. (a), (b), and (c) show, the uncertainty 
$\Delta y \Delta p_y$ in graphene exhibits a similar behavior to $\Delta x \Delta p_x$. However, the critical values $\mu_1$ and $\mu_2$ are
changed into $\nu_1$ and $\nu_2$. (d) The critical value 
$\nu_2$ increases with decreasing $\beta$, and eventually goes to $\infty$ at $\beta = 0$.}
\end{center}
\end{figure}

The dimensionless uncertainty $\Delta y \Delta p_y / \hbar$ is plotted in Fig. 2 when $a=0.9$, $d = 8 (nm)$, $\alpha = 1.2 (1/nm)$ and 
$\beta = 0.04 (1/nm)$. We also choose $\lambda_c^{-1}$ as $8 (1/nm)$ (Fig. 2a), $2 (1 / nm)$ (Fig. 2b), and $0.08 (1/nm)$ (Fig. 2c).
We plot $(\Delta y \Delta p_y / \hbar)_{free}$ together for comparison. As Fig. 2 shows, $\Delta y \Delta p_y$ exhibits a similar behavior 
with $\Delta x \Delta p_x$. However, the critical values $\mu_1$ and $\mu_2$ are changed into $\nu_1$ and $\nu_2$, which reduce to
\begin{equation}
\label{critical3}
\nu_1 = \frac{1}{\sqrt{2} d},   \hspace{1.5cm}   \delta (\lambda_c^{-1}) \bigg|_{\lambda_c^{-1} = \mu_2} = 1,
\end{equation}
where
\begin{equation}
\label{critical4}
\delta (\lambda_c^{-1}) = \frac{2 \lambda_c^{-2} d^4}{\pi} \left[J_{0,2} - \frac{d^2}{\pi}\left\{ (a^2 - b^2) \lambda_c^{-1} J_{0,1} + 2 a b J_{1,1} \right\}^2 \right].
\end{equation}
The $\lambda_c^{-1}$-dependence of $\delta (\lambda_c^{-1})$ is plotted in Fig. 2d when $a = 0.9$, $d = 8 (nm)$, $\alpha = 1.2 (1/nm)$, 
and $\beta = 1.2 / n (1/nm)$ for various $n$. As this figure indicates, the critical value $\nu_2$ increases with increasing $n$, and eventually 
goes to $\infty$ when $\beta = 0$.

\section{position-velocity uncertainty}

In this section we discuss on the position-velocity uncertainties \cite{po-vel}, which is completely different from position-momentum 
uncertainties because of ${\bm p} \neq M {\bm v}$. The velocity operator 
$\hat{v}_x (t)$ is defined as $\exp \left(i \hat{H}_M t / \hbar \right) \hat{v}_x (0) \exp \left(-i \hat{H}_M t / \hbar \right)$, where 
$\hat{v}_x (0) = \partial \hat{H}_M / \partial p_1$. This operator is easily constructed from $\hat{x} (t)$ by making use of 
Ehrenfest\cite{schiff} theorem $d \hat{x} (t) / d t = (i / \hbar) \exp \left(i \hat{H}_M t / \hbar \right) [\hat{H}_M, \hat{x} (0)] \exp \left(-i \hat{H}_M t / \hbar \right) = \hat{v}_x (t)$.  Then, the final expression of $\hat{v}_x (t)$ is 
\begin{eqnarray}
\label{velocity1}
\hat{v}_x (t) = \left(             \begin{array}{cc}
                       \hat{U} (p)  &  \hat{u}_1 (p) + i \hat{u}_2 (p)    \\
                       \hat{u}_1 (p) - i \hat{u}_2 (p)  &  -\hat{U} (p)
                                   \end{array}                            \right),
\end{eqnarray}
where
\begin{eqnarray}
\label{velocity2}
& &\hat{U} (p) = v_F \left[ \frac{2 p_2}{\sqrt{{\bm p}^2 + (M v_F)^2}} \sin \theta_M \cos \theta_M + 
\frac{2 (M v_F) p_1}{{\bm p}^2 + (M v_F)^2} \sin^2 \theta_M \right]                  \\   \nonumber
& &\hat{u}_1 (p) = v_F \left[ \cos^2 \theta_M + \frac{p_1^2 - p_2^2 - (M v_F)^2}{{\bm p}^2 + (M v_F)^2} \sin^2 \theta_M \right]
                                                                                     \\   \nonumber
& &\hat{u}_2 (p) = v_F \left[- \frac{2 p_1 p_2}{{\bm p}^2 + (M v_F)^2} \sin^2 \theta_M + \frac{2 (M v_F)}{\sqrt{{\bm p}^2 + (M v_F)^2}}
                              \sin \theta_M \cos \theta_M  \right].
\end{eqnarray}
Unlike the position operators $\hat{x} (t)$ and $\hat{y} (t)$ the velocity operator $\hat{v}_x (t)$ does not have the spreading 
term. This is due to the fact that the spreading term in the position operators is linear in time. Another remarkable property of 
$\hat{v}_x (t)$ is that $\hat{v}_x^2 (t)$ is simply $v_F^2$ times identity operator $\openone$. Combining these two properties one can easily 
conjecture $\lim_{t \rightarrow \infty} \Delta v_x = v_F$ regardless of the choice of the wave packet because the ZB term in 
$\hat{v}_x (t)$ has infinitely high frequency in this limit, and therefore, is canceled out in the time average.

\begin{figure}[ht!]
\begin{center}
\includegraphics[height=6.0cm]{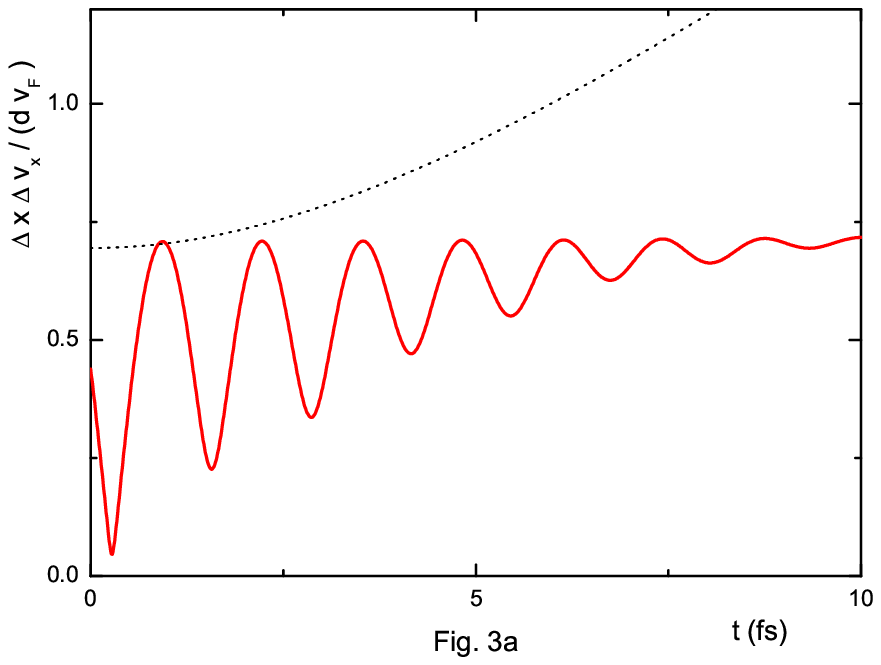}
\includegraphics[height=6.0cm]{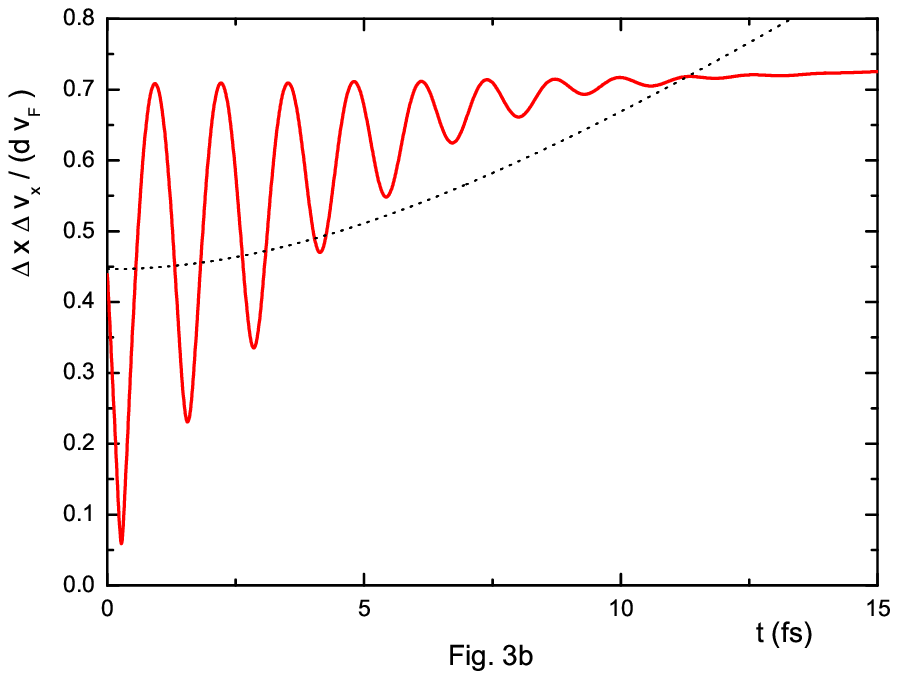}
\includegraphics[height=6.0cm]{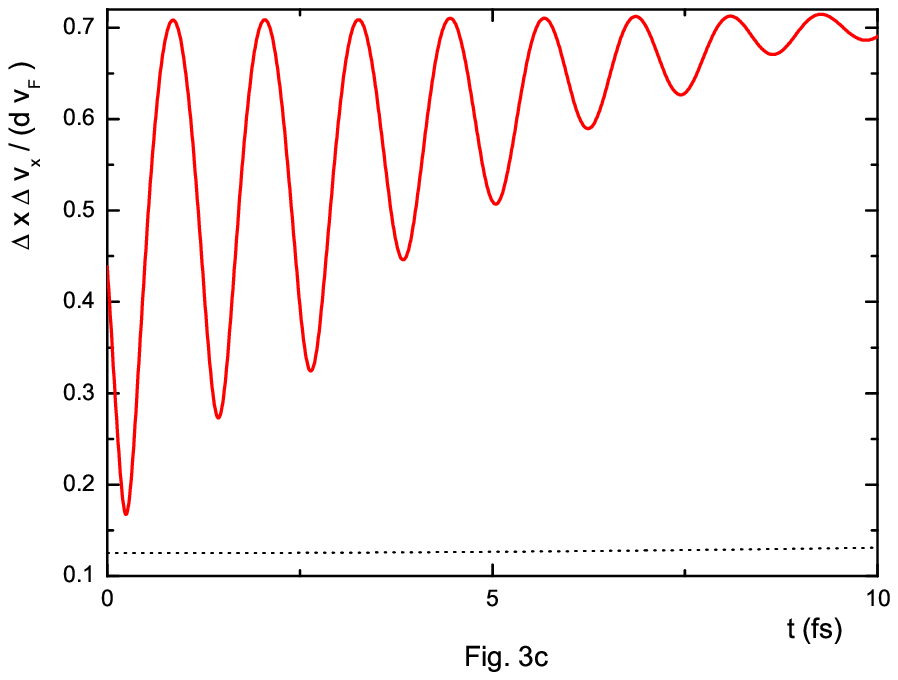}
\caption[fig3]{(Color online) The time-dependence of $\Delta x \Delta v_x / d v_F$ for $\lambda_c^{-1} = 0.09 (1 / nm)$ (a), 
$\lambda_c^{-1} = 0.14 (1 / nm)$ (b), 
and $\lambda_c^{-1} = 0.5 (1 / nm)$ (c). The black dotted line for each figure is a corresponding value $(\Delta x \Delta v_x / d v_F)_{free}$. 
As Fig. (a), (b), and (c) show, the uncertainty 
$\Delta x \Delta v_x$ in graphene is larger (or smaller) than $(\Delta x \Delta v_x)_{free}$ depending on the gap parameter $\lambda_c^{-1}$.
One can show explicitly that  $\lim_{t \rightarrow 0} \Delta x \Delta v_x < (\Delta x \Delta v_x)_{free}$ if $\lambda_c^{-1} < \mu_1$ and $\lim_{t \rightarrow \infty} \Delta x \Delta v_x > (\Delta x \Delta v_x)_{free}$ if 
$\lambda_c^{-1} > \mu_{2*}$, where $\mu_{2*}$ is defined as $\gamma (\lambda_c^{-1} = \mu_{2*}) = 1 / (2 (\mu_{2 *} d)^2$. }
\end{center}
\end{figure}

The expectation value $\langle v_x \rangle (t)$ and $\langle v_x^2 \rangle (t)$ with a wave packet (\ref{packet}) can be straightforwardly 
computed by making use of Eq. (\ref{velocity1}). As expected the resulting $\Delta v_x (t)$ has only trembling motion and approaches to 
$v_F$ at $t \rightarrow \infty$ limit. The dimensionless position-velocity uncertainty $\Delta x \Delta v_x / d v_F$ is plotted in Fig. 3 for 
$\lambda_c^{-1} = 0.09 (1 / nm)$ (Fig. 3a), $\lambda_c^{-1} = 0.14 (1 / nm)$ (Fig. 3b), and $\lambda_c^{-1} = 0.5 (1 / nm)$ (Fig. 3c) when 
$a = 0.9$, $d = 8 (nm)$, $\alpha = 0.04 (1 / nm)$ and $\beta = 1.2 (1 / nm)$. The $x$-axis is time axis with femto-second unit. The black dotted
line is a corresponding value $(\Delta x \Delta v_x)_{free} / d v_F$, where 
$(\Delta x \Delta v_x)_{free} = \sqrt{\lambda_c^2 v_F^2 / 4 + \lambda_c^4 v_F^4 t^2 / 4 d^4}$ is a position-velocity uncertainty for $\hat{H}_{free}$. The overall increasing behavior of $\Delta x \Delta v_x$ is solely due to $\Delta x$ 
because $\Delta v_x$ does not have its own spreading term. As Fig. 3 shows, $\Delta x \Delta v_x$ can be smaller or larger than 
$(\Delta x \Delta v_x)_{free}$ depending on the gap parameter $\lambda_c$. In order to compare $\Delta x \Delta v_x$ with 
$(\Delta x \Delta v_x)_{free}$ more accurately we compute its limiting values at $t \rightarrow 0$ and $t \rightarrow \infty$. Then, 
it is easy to show $\lim_{t \rightarrow 0} \Delta x \Delta v_x < (\Delta x \Delta v_x)_{free}$ if $\lambda_c^{-1} < \mu_1$, where $\mu_1$ is 
defined at Eq.(\ref{critical1}), and $\lim_{t \rightarrow \infty} \Delta x \Delta v_x > (\Delta x \Delta v_x)_{free}$ if 
$\lambda_c^{-1} > \mu_{2*}$, where $\mu_{2*}$ is defined as $\gamma (\lambda_c^{-1} = \mu_{2*}) = 1 / (2 (\mu_{2 *} d)^2$. 
The critical values $\mu_1$, $\mu_2$, and $\mu_{2*}$ are given at Table I when $d = 8 (nm)$, $\alpha = 1.2 / n (1 / nm)$, $\beta = 1.2 (1 / nm)$,
and $a = 0.9$ or $0.7$. The reason for choice of $a$ is that while the diagonal components of the various operators contribute
dominantly to the uncertainty relations at $a = 0.9 \sim 1$, the off-diagonal components become more important at $a = 0.7 \sim 1 / \sqrt{2}$. 
As expected from Fig. 1d, $\mu_2$ increases with increasing $n$, and eventually goes to $\infty$ at $\alpha = 0$. Another critical value
$\mu_{2*}$ also exhibits an increasing behavior with increasing $n$, but its increasing rate is very small compared to $\mu_2$ and 
converges to $0.332$ at $n \rightarrow \infty$ limit.

\begin{center}
{\large{Table I}}: Critical values for $\Delta x \Delta p_x$ and $\Delta x \Delta v_x$ when $ d = 8 (nm)$, $\alpha = 1.2 / n (1 / nm)$ and 
$\beta = 1.2 (1 / nm)$.
\end{center}

\begin{center}
\begin{tabular}{c|c|cccccc}  \hline  \hline
 & $a$ & $n=10$ & $n=20$ & $n=30$ & $n=40$ & $n=50$ & $n=\infty$         \\  \hline
 $\mu_1 (1 / (nm)$  &  $0.9$  &  $0.143$  &  $0.143$  & $0.143$  &  $0.143$  & $0.143$  &  $0.143$  \\  \cline{2-8}
 & $0.7$  &  $4.42$  &  $4.42$  &  $4.42$  &  $4.42$  &  $4.42$  &  $4.42$   \\   \hline
 $\mu_2 (1 / (nm)$  &  $0.9$  &  $1.03$  &  $2.24$  & $3.47$  &  $4.69$  & $5.90$  &  $\infty$  \\  \cline{2-8}
 & $0.7$  &  $0.90$  &  $1.79$  &  $2.68$  &  $3.58$  &  $4.47$  &  $\infty$   \\   \hline  
$\mu_{2*} (1 / (nm)$  &  $0.9$  &  $0.257$  &  $0.303$  & $0.318$  &  $0.324$  & $0.327$  &  $0.332$  \\  \cline{2-8}
 & $0.7$  &  $0.256$  &  $0.302$  &  $0.317$  &  $0.323$  &  $0.326$  &  $0.332$   \\   \hline  \hline
\end{tabular}
\\
\end{center}
\vspace{0.5cm}


Following similar calculation procedure one can plot the time-dependence of the dimensionless quantity $\Delta y \Delta v_y / (d v_F)$. 
Although the time-dependence of the uncertainties is not plotted in this paper, $\Delta y \Delta v_y$ exhibits a similar behavior with 
$\Delta x \Delta v_x$. However, the critical values $\mu_1$ and $\mu_{2*}$ are changed into $\nu_1$ and $\nu_{2*}$, whose explicit values
are given at Table II.

\begin{center}
{\large{Table II}}: Critical values for $\Delta y \Delta p_y$ and $\Delta y \Delta v_y$ when $ d = 8 (nm)$, $\alpha = 1.2 (1 / nm)$ and 
$\beta = 1.2 / n (1 / nm)$.
\end{center}

\begin{center}
\begin{tabular}{c|c|cccccc}  \hline  \hline
 & $a$ & $n=10$ & $n=20$ & $n=30$ & $n=40$ & $n=50$ & $n=\infty$         \\  \hline
 $\nu_1 (1 / (nm)$  &  $0.9$  &  $0.088$  &  $0.088$  & $0.088$  &  $0.088$  & $0.088$  &  $0.088$  \\  \cline{2-8}
 & $0.7$  &  $0.088$  &  $0.088$  &  $0.088$  &  $0.088$  &  $0.088$  &  $0.088$   \\   \hline
 $\nu_2 (1 / (nm)$  &  $0.9$  &  $2.23$  &  $3.36$  & $4.48$  &  $5.60$  & $6.73$  &  $\infty$  \\  \cline{2-8}
 & $0.7$  &  $1.22$  &  $2.05$  &  $2.88$  &  $3.73$  &  $4.59$  &  $\infty$   \\   \hline  
$\nu_{2*} (1 / (nm)$  &  $0.9$  &  $0.309$  &  $0.326$  & $0.329$  &  $0.330$  & $0.331$  &  $0.332$  \\  \cline{2-8}
 & $0.7$  &  $0.319$  &  $0.328$  &  $0.330$  &  $0.331$  &  $0.331$  &  $0.332$   \\   \hline  \hline
\end{tabular}
\\
\end{center}
\vspace{0.5cm}

\section{concluding Remarks}

In this paper we have examined the position-momentum and position-velocity uncertainties for the monolayer gapped graphene. We have shown that the uncertainties are contributed by the spreading effect of the wave packet in the long-range of time and the ZB in the short-range of time. By choosing the gap parameter $\lambda_c$ appropriately one can control the uncertainties within the quantum mechanical law. 

The uncertainties can be tested experimentally because all figures in this paper show a significant difference between free and graphene cases. The uncertainties in the graphene might be measured via the following one-slit experiment (see Figure $4$). In this paper we will discuss on $\Delta x$ only because other quantities can be measured similarly. The slit width $d$ should be order of Angstroms to ensure the occurrence of diffraction in the slit. The distance $L$ should be order of nanometers because the effect of the zitterbewegung is important within initial few femtoseconds. The electrons emitted by the emitter would arrive at the detecter through the slit. Then, one can make a probability distribution with respect to $x$, which would be a smooth Gaussian form. Measuring the width of the Gaussian distribution, one can deduce $\Delta x$ at $t \sim L / v_F$, where $v_F$ is a Fermi velocity. Repeating the same experiment with changing $L$ one can measure the time-dependence of $\Delta x$. If the prediction we presented in this paper is correct, $\Delta x$ would exhibit an oscillating behavior in the short-range of time due to the effect of the zitterbewegung, but globally an increasing behavior in the long-range of time due to the spreading effect of the wave packet. 

\begin{figure}[ht!]
\begin{center}
\includegraphics[height=6cm]{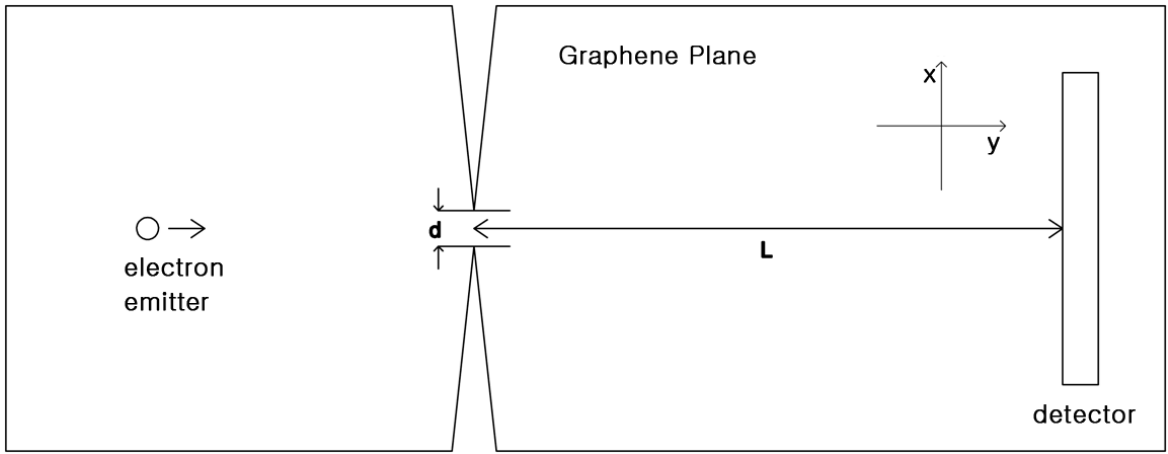}
\caption[fig3]{(Color online) Schematic diagram for measuring the uncertainties.}
\end{center}
\end{figure}


It is interesting to extend this paper to the bilayer graphene. Another interesting issue is to examine the uncertainty relations when 
external magnetic field is applied. We guess that the external magnetic field drastically reduce the uncertainties in the graphene. If so, 
the graphene-based quantum computer can be more useful for huge calculations. We would like to explore this issue in the near future.

{\bf Acknowledgement}:
This research was supported by Basic Science Research Program through the National Research Foundation of Korea(NRF) funded by the Ministry of Education, Science and Technology(2011-0011971 and National Scientist Program 2010-0020414).

\newpage

\begin{appendix}{\centerline{\bf Appendix A}}

\setcounter{equation}{0}
\renewcommand{\theequation}{A.\arabic{equation}}
In this appendix we summarize the various expectation values at $\alpha = 0$ and $a = b = 1 / \sqrt{2}$, where Eq. (\ref{decompose3}) and Eq. (\ref{decompose4}) imply that the 
initial wave packet has equal intensity of positive-energy and negative-energy states. In this simple case the expectation values 
$\langle x \rangle (t)$ and $\langle y \rangle (t)$ reduce to 
\begin{eqnarray}
\label{appen-b-1}
& &\langle x \rangle (t) = \frac{d^2}{\pi} \int d^2 {\bf k} e^{-d^2 k_x^2 - d^2 (k_y - \beta)^2}
\left[ (v_F t) \frac{k_x^2}{{\bf k}^2 + \lambda_c^{-2}} + \sin \theta \cos \theta \frac{k_y^2 + \lambda_c^{-2}}{({\bf k}^2 + \lambda_c^{-2})^{3/2}}
                                                                                               \right]
                                                                                                                                       \\   \nonumber
& &\langle y \rangle (t) = \frac{d^2 \lambda_c^{-1}}{\pi} \int d^2 {\bf k} e^{-d^2 k_x^2 - d^2 (k_y - \beta)^2}
\frac{\sin^2 \theta}{{\bf k}^2 + \lambda_c^{-2}}
\end{eqnarray}
where $\theta = (v_F t) \sqrt{{\bf k}^2 + \lambda_c^{-2}}$. In the case of zero gap we get $\langle y \rangle (t) = 0$. Since 
$\langle x^2 \rangle (t)$ and $\langle y^2 \rangle (t)$ are independent of choice of $a$ and $b$, those are equal to 
Eq. (\ref{psquare2}) and Eq. (\ref{psquare4}) with $\alpha = 0$. The expectation values for the velocity operators becomes
\begin{eqnarray}
\label{appen-b-2}
& &\langle v_x \rangle (t) = v_F - \frac{2 v_F d^2}{\pi}
\int d^2 {\bf k} e^{-d^2 k_x^2 - d^2 (k_y - \beta)^2} \sin^2 \theta \frac{k_y^2 + \lambda_c^{-2}}{{\bf k}^2 + \lambda_c^{-2}}         \\   \nonumber
& &\langle v_y \rangle (t) = \frac{v_F d^2 \lambda_c^{-1}}{\pi}
\int d^2 {\bf k} e^{-d^2 k_x^2 - d^2 (k_y - \beta)^2} \frac{\sin 2 \theta}{\sqrt{{\bf k}^2 + \lambda_c^{-2}}}.
\end{eqnarray}
In the case of zero gap we also get $\langle v_y \rangle (t) = 0$. Of course, expectation values for the square of velocity operators are 
simply $\langle v_x^2 \rangle = \langle v_y^2 \rangle = v_F^2$.
\end{appendix}

\newpage

\begin{appendix}{\centerline{\bf Appendix B}}

\setcounter{equation}{0}
\renewcommand{\theequation}{B.\arabic{equation}}
In this appendix we summarize the explicit expressions for $\langle x \rangle (t)$, $\langle y \rangle (t)$, $\langle x^2 \rangle (t)$, 
$\langle y^2 \rangle (t)$, $\langle v_x \rangle (t)$, and $\langle v_y \rangle (t)$ by making use of the binomial expansion and performing
the ${\bf k}$-integration. The integral formula we use is 
\begin{equation}
\label{appen-a-1}
\int_{-\infty}^{\infty} x^n e^{-(x - \beta)^2} dx = (2 i)^{-n} \sqrt{\pi} H_n (i \beta),
\end{equation}
where $H_n(z)$ is the usual Hermite polynomial.

The expectation values $\langle x \rangle (t)$ and $\langle y \rangle (t)$ expressed in Eq. (\ref{position3}) and Eq. (\ref{position5}) 
reduce to 
\begin{eqnarray}
\label{appen-a-2}
& &\langle x \rangle (t) = 2ab v_F t + \sum_{n=0}^{\infty} \frac{(2 \lambda_c^{-1} v_F t)^{2n+2}}{(2n + 3)!}
     \sum_{\ell = 0}^n  \left(  \begin{array}{c} n \\ \ell   \end{array} \right) \frac{(-1)^{n - \ell}}{(2 \lambda_c^{-1} d)^{2 \ell + 2}}
                                                                                                                              \\   \nonumber
& &\hspace{4.0cm} \times
     \sum_{m=0}^{\ell} \left(  \begin{array}{c} \ell \\ m   \end{array} \right)
     \bigg[-i (a^2 - b^2) d X_1 + 2 a b (v_F t) X_2 \bigg]                                                                    \\   \nonumber
& &\langle y \rangle (t) = \sum_{n=0}^{\infty} \frac{(2 \lambda_c^{-1} v_F t)^{2n+2}}{(2n + 3)!}
     \sum_{\ell = 0}^n  \left(  \begin{array}{c} n \\ \ell   \end{array} \right) \frac{(-1)^{n - \ell}}{(2 \lambda_c^{-1} d)^{2 \ell + 2}}
     \sum_{m=0}^{\ell} \left(  \begin{array}{c} \ell \\ m   \end{array} \right)
     \bigg[i (a^2 - b^2) d Y_1 +  a b \lambda_c Y_2 \bigg] 
\end{eqnarray}
where
\begin{eqnarray}
\label{appen-a-3}
& &X_1 = (2n+3) H_{2m} (i\alpha d) H_{2\ell -2m+1}(i \beta d) + 2(\lambda_c^{-1} v_F t)H_{2m+1}(i\alpha d) H_{2\ell-2m}(i \beta d)
                                                                                                                                \\   \nonumber
& &X_2 =  H_{2m} (i\alpha d) H_{2\ell -2m+2}(i \beta d) - (2 \lambda_c^{-1} d)^2 H_{2m}(i\alpha d) H_{2\ell-2m}(i \beta d)      \\   \nonumber
& &Y_1 = (2n+3) H_{2m+1} (i\alpha d) H_{2\ell -2m}(i \beta d) + 2(\lambda_c^{-1} v_F t)H_{2m}(i\alpha d) H_{2\ell-2m+1}(i \beta d)
                                                                                                                                 \\   \nonumber
& &Y_2 = (2n+3)(2 \lambda_c^{-1} d)^2 H_{2m} (i\alpha d) H_{2\ell -2m}(i \beta d) - 2(\lambda_c^{-1} v_F t)H_{2m+1}(i\alpha d) 
                                                                                                             H_{2\ell-2m+1}(i \beta d).
\end{eqnarray}
Although the arguments of the Hermite polynomials are pure imaginary, one can show easily that $\langle x \rangle (t)$ and $\langle y \rangle (t)$
are real by considering the fact that $H_n(z)$ is even (or odd) function when $n$ is even (or odd).

Similarly one can express $\langle x^2 \rangle (t)$ and $\langle y^2 \rangle (t)$ from Eq. (\ref{psquare2}) and Eq. (\ref{psquare4}) as follows:
\begin{eqnarray}
\label{appen-a-4}
& &\langle x^2 \rangle (t) = \frac{d^2}{2} + (v_F t)^2                                                                           \\   \nonumber
& &\hspace{1.0cm}
+ 2 d^2 \sum_{n=0}^{\infty} \frac{(2 \lambda_c^{-1} v_F t)^{2n+4}}{(2n + 4)!}
     \sum_{\ell = 0}^n  \left(  \begin{array}{c} n \\ \ell   \end{array} \right) \frac{(-1)^{n - \ell}}{(2 \lambda_c^{-1} d)^{2 \ell + 4}}
     \sum_{m=0}^{\ell} \left(  \begin{array}{c} \ell \\ m   \end{array} \right) X_3                                               \\   \nonumber
& &\langle y^2 \rangle (t) = \frac{d^2}{2} + (v_F t)^2                                                                           \\   \nonumber
& &\hspace{1.0cm}
+ 2 d^2 \sum_{n=0}^{\infty} \frac{(2 \lambda_c^{-1} v_F t)^{2n+4}}{(2n + 4)!}
     \sum_{\ell = 0}^n  \left(  \begin{array}{c} n \\ \ell   \end{array} \right) \frac{(-1)^{n - \ell}}{(2 \lambda_c^{-1} d)^{2 \ell + 4}}
     \sum_{m=0}^{\ell} \left(  \begin{array}{c} \ell \\ m   \end{array} \right) Y_3
\end{eqnarray}
where $X_3 = X_2$ and 
\begin{equation}
\label{appen-a-5}
Y_3 =  H_{2m+2} (i\alpha d) H_{2\ell -2m}(i \beta d) - (2 \lambda_c^{-1} d)^2 H_{2m}(i\alpha d) H_{2\ell-2m}(i \beta d).
\end{equation}

Although we have not derived the integral representations of $\langle v_x \rangle (t)$ and $\langle v_y \rangle (t)$ explicitly in the 
main text, their derivations are straightforward. Then, the expressions of $\langle v_x \rangle (t)$ and $\langle v_y \rangle (t)$ in terms 
of the Hermite polynomials are 
\begin{eqnarray}
\label{appen-a-6}
& &\langle v_x \rangle (t) = 2ab v_F - 2 v_F \sum_{n=0}^{\infty} \frac{(2 \lambda_c^{-1} v_F t)^{2n+1}}{(2n + 2)!}
     \sum_{\ell = 0}^n  \left(  \begin{array}{c} n \\ \ell   \end{array} \right) \frac{(-1)^{n - \ell}}{(2 \lambda_c^{-1} d)^{2 \ell + 2}}
                                                                                                                                      \\  \nonumber
& &\hspace{3.5cm} \times
     \sum_{m=0}^{\ell} \left(  \begin{array}{c} \ell \\ m   \end{array} \right)
     \bigg[i (a^2 - b^2) (2 \lambda_c^{-1} d) U_1 + 2 a b (\lambda_c^{-1} v_F t) U_2 \bigg]                                           \\  \nonumber
& &\langle v_y \rangle (t) = v_F \sum_{n=0}^{\infty} \frac{(2 \lambda_c^{-1} v_F t)^{2n+1}}{(2n + 2)!}
     \sum_{\ell = 0}^n  \left(  \begin{array}{c} n \\ \ell   \end{array} \right) \frac{(-1)^{n - \ell}}{(2 \lambda_c^{-1} d)^{2 \ell + 1}}
     \sum_{m=0}^{\ell} \left(  \begin{array}{c} \ell \\ m   \end{array} \right)
     \bigg[i (a^2 - b^2) V_1 + 2 a b V_2 \bigg]  
\end{eqnarray}
where
\begin{eqnarray}
\label{appen-a-7}
& &U_1 = (n+1) H_{2m} (i\alpha d) H_{2\ell -2m+1}(i \beta d) + (\lambda_c^{-1} v_F t)H_{2m+1}(i\alpha d) H_{2\ell-2m}(i \beta d)          \\  \nonumber
& &U_2 = (2 \lambda_c^{-1} d)^2 H_{2m} (i\alpha d) H_{2\ell -2m}(i \beta d) - H_{2m}(i\alpha d) H_{2\ell-2m+2}(i \beta d)                 \\  \nonumber
& &V_1 = (2n+2) H_{2m+1} (i\alpha d) H_{2\ell -2m}(i \beta d) - 2 (\lambda_c^{-1} v_F t)H_{2m}(i\alpha d) H_{2\ell-2m+1}(i \beta d)       \\  \nonumber
& &V_2 = (2n+2) (2 \lambda_c^{-1} d) H_{2m} (i\alpha d) H_{2\ell -2m}(i \beta d) -  \frac{v_F t}{d} H_{2m+1}(i\alpha d) H_{2\ell-2m+1}(i \beta d).
\end{eqnarray}
\end{appendix}


\begin{thebibliography}{99}
\bibitem{fabrication} K. Novoselov, A. K. Geim, S. V. Morozov, D. Jiang, M. I. Katsnelson, I. V. Grigorieva, S. V. Dubonos, and A. A. Firsov, 
Nature (London) {\bf 438} (2005) 197; K. S. Kim et al., Nature {\bf 457} (2009) 706; S. Bae et al., Nat. Nanotechnol. {\bf 5} (2010) 574.
\bibitem{review} For a recent review, see A. H. Castro Neto, F. Guinea, N. M. R. Peres, K. S. Novoselov, and A. K. Geim, Rev. Mod. Phys. {\bf 81} (2009) 109 and references therein.
\bibitem{wallace} P.~R.~Wallace, Phys. Rev. {\bf 71} (1947) 622.
\bibitem{semenoff} G.~W.~Semenoff, Phys. Rev. Lett. {\bf 53} (1984) 2449.
\bibitem{connection} R. Jackiw and S. -Y. Pi, Phys. Rev. Lett. {\bf 98} (2007) 266402; J. K. Pachos and M. Stone, Int. J. Mod. Phys. {\bf B 21} (2007) 5113; D. Allor, T. D. Cohen, and D. A. McGady, Phys. Rev. {\bf D 78} (2008) 096009; C. G. Beneventano and E. M. Santangelo, J. Phys.{\bf A 41} (2008) 164035; R. Jackiw and S. -Y. Pi, arXiv:0808.1562 (cond-mat); E. M. Santangelo. arXiv:0809.4844 (hep-th); C. G. Beneventano, P. Giacconi,
E. M. Santangelo, and R. Soldati, J. Phys. {\bf A 42} (2009) 275401; M. Bordag, I. V. Fialkovsky, D. M. Gitman, and D. V. Vassilevich, Phys. Rev. {\bf B 80} (2009) 245406; G. W. Semenoff, arXiv:1005.0572 (hep-th); A. Iorio, Ann. Phys. {\bf 326} (2011) 1334; L. B. Drissi, E. H. Saidi, and M. Bousmina,
Nucl. Phys. {\bf B 829} (2010) 523. 
\bibitem{ab1959} Y. Aharonov and D. Bohm, Phys. Rev. {\bf 115} (1959) 485.
\bibitem{spin-ab} R. Jackiw, ``{\it Delta function potential in two- and three-dimensional quantum mechanics}'', in M. A. B. Beg memorial volume, A. Ali and P. Hoodbhoy, eds., (World Scientific, Singapore, 1991); Ph. de Sousa Gerbert, Phys. Rev. {\bf D 40} (1989) 1346; C. R. Hagen,
Phys. Rev. Lett. {\bf 64} (1990) 503; D. K. Park, J. Math. Phys. {\bf 36} (1995) 5453; D.K.Park and S. -K. Yoo, Ann. Phys. {\bf 263} (1998) 295.
\bibitem{cosmic} M. G. Alford and F. Wilczek, Phys. Rev. Lett. {\bf 62} (1989) 1071; M. G. Alford, J. March-Russell and F. Wilczek,
Nucl. Phys. {\bf B328} (1989) 140; Y. H. Chen, F. Wilczek, E. Witten and B. I. Halperin, Int. J. Mod. Phys. {\bf B3} (1989) 1001.
\bibitem{ab-graphene-th} P. Recher, B. Trauzettel, A. Rycerz, Y. M. Blater, C. W. J. Beenakker, and A. F. Morpurgo, Phys. Rev. {\bf B 76} (2007) 235404; 
 A. Rycerz, Acta. Phys. Pol. {\bf A 115} (1009) 322; J. Wurm, M. Wimmer, H. U. Baranger, and K. Richter, Semicond. Sci. Technol. {\bf 25} (2010) 034003; 
 A. Rycerz and C. W. J. Beenakker, arXiv: 0709.3397 (cond-mat); R. Jackiw, A. I. Milstein, S. -Y. Pi, and I. S. Terekhov, Phys. Rev. {\bf B 80} (2009) 033413; M. I. Katsnelson, Europhys. Lett. {\bf 89} (2010) 17001; J. Schelter, D. Bohr and B. Trauzettel, Phys. Rev. {\bf B 81} (2010) 195441. 
\bibitem{ab-graphene-exp} S. Russo {\it et al}, Phys. Rev. {\bf B 77} (2008) 085413; F. Molitor {\it et al},  New J. Phys. {\bf 12} (2010) 043054. 
\bibitem{coulomb} V. M. Pereira, J. Nilsson, and A. H. Castro Neto, Phys. Rev. Lett. {\bf 99} (2007) 166802; A. V. Shytov, M. I. Katsnelson,
and L. S. Levitov, Phys. Rev. Lett. {\bf 99} (2007) 236801; A. V. Shytov, M. I. Katsnelson, and L. S. Levitov, Phys. Rev. Lett. {\bf 99} (2007) 246802;
V. M. Pereira, V. N. Kotov, A. H. Castro Neto,  Phys. Rev. {\bf B 78} (2008) 085101;
J. Wang, H.A. Fertig and G. Murthy, Phys. Rev. Lett. {\bf 104} (2010) 186401; F. de Juan, A. G. Grushin and  M. A. H. Vozmediano, Phys. Rev. {\bf B 82} (2010) 125409. 
\bibitem{zeldo} Y. B. Zeldovich and V. S. Popov, Sov. Phys. Usp. {\bf 14} (1972) 673.
\bibitem{klein} O. Klein, Z. Phys. {\bf 53} (1929) 157; N. Dombey and A. Calogeracos, Phys. Rep. {\bf 315} (1999) 41.
\bibitem{paradox1} M. I. Katsnelson, K. Novoselov, and A. K. Geim,  
Nature (London) {\bf 2} (2006) 620.
\bibitem{paradox2} A. F. Young and P. Kim, Nature Physics {\bf 5} (2009) 222.
\bibitem{zitter1} J. D. Bjorken and S. D. Drell, {\it Relativistic Quantum Mechanics} (McGraw-Hill, New York, 1964).
\bibitem{zitter2} T. M. Rusin and W. Zawadzki, Phys. Rev. {\bf B 76} (2007) 195439; 
J. Schliemann, New J. Phys. {\bf 10} (2008) 043024;
G. M. Maksimova, V. Ya. Demikhovskii, and E. V. Frolova, Phys. Rev. {\bf B 78} (2008) 235321;
\bibitem{zitter3} T. M. Rusin and W. Zawadzki, Phys. Rev. {\bf B 78} (2008) 125419;
E. Romera and F. de los Santos, Phys. Rev. {\bf B 80} (2009) 165416;
W. Zawadzki and T. M. Rusin, J. Phys.: Cond. Matt. {\bf 23} (2011) 143201.
\bibitem{zitter4} V. Ya. Demikhovskii, G. M. Maksimova, and E. V. Frolova, arXiv:0912.0331 (cond-mat);
V. Ya. Demikhovskii, G. M. Maksimova, A. A. Perov, and E. V. Frolova, arXiv:1007.1566 (quant-ph).
\bibitem{qcs} T. D. Ladd {\it et al}, Nature {\bf 464} (2010) 45 and references therein.
\bibitem{qc1} B. Trauzettel, D. V. Bulaev, D. Loss and G. Burkard, Nature Physics {\bf 3} (2007) 192; 
P. Recher, J. Nilsson, G. Burkard, and B. Trauzettel, Phys. Rev. {\bf B 79} (2009) 085407;
P. Recher and B. Trauzettel, Nanotechnology {\bf 21} (2010) 302001;
T. Dirks {\it et al}, Nature Physics {\bf 7} (2011) 386;
G. Y. Wu, N.-Y. Lue and L. Chang, arXiv:1104.0443 (cond-mat). 
\bibitem{teleport} C. H. Bennett {\it et al}, Phys. Rev. Lett. 70 (1993) 1895.
\bibitem{factoring} P. W. Shor, Algorithms for Quantum Computation: Discrete Logarithms and Factoring, Proc.
35th Annual Symposium on Foundations of Computer Science (1994) 124.
\bibitem{search} L. K. Grover, Phys. Rev. Lett. 79 (1997) 325.
\bibitem{dynamical3} Y. Araki and T. Hatsuda, Phys. Rev. {\bf B 82} (2010) 121403(R).
\bibitem{DGSB} E. Farhi and R. Jackiw, {\it Dynamical Gauge Symmetry Breaking} (World Scientific, Singapore, 1982),
\bibitem{dynamical1} S. Y. Zhou, G. -H. Gweon, A. V. Fedorov, P. N. First, W. A. de Heer, D. -H. Lee, F. Guinea, A. H. Castro Neto, and A. Lanzara, 
Nature Material {\bf 6} (2007) 770.
\bibitem{dynamical2} M. Y. Han, B. \"{O}zyilmaz, Y. Zhang, and P. Kim, Phys. Rev. Lett. {\bf 98} (2007) 206805; 
W. Y. Kim and K. S. Kim, Nat. Nanotechnol. {\bf 3} (2008) 408.
\bibitem{bilayer} E. McCann and V. I. Fal'ko, Phys. Rev. Lett. {\bf 96} (2006) 086805.
\bibitem{po-vel} M. H. Al-Hashimi and U. -J. Wiese, arXiv:0907.5178 (quant-ph).
\bibitem{schiff} L. L. Schiff, {\it Quantum Mechanics} (McGraw-Hill, Tokyo, 1968).




\end{thebibliography}
\end{document}